\documentclass[]{elsarticle}
\pdfoutput=1
\biboptions{sort&compress}
\usepackage[english]{babel}
\usepackage[utf8x]{inputenc}
\usepackage{amsmath}
\usepackage{amssymb}
\usepackage{wasysym}
\usepackage{graphicx}
\usepackage{color}
\usepackage{ulem}
\usepackage[colorinlistoftodos]{todonotes}
\usepackage{lineno}
\usepackage{hyperref}
\usepackage[shortlabels]{enumitem}
\usepackage{setspace}
\usepackage{verbatim}

\begin{document}
\begin{frontmatter}

\title{Sensitivity of alkali halide scintillating calorimeters with particle identification to investigate the DAMA dark matter detection claim}

\author[qu]{P.~Nadeau\corref{cor1}}
\ead{pnadeau@owl.phy.queensu.ca}
\author[qu]{M.~Clark}
\author[qu]{P.~C.~F.~Di Stefano}
\author[tum]{J.-C.~Lanfranchi}
\author[tum]{S.~Roth}
\author[tum]{M.~von~Sivers\fnref{fn1}}
\author[pi,mac]{I.~Yavin}

\cortext[cor1]{Corresponding author}

\fntext[fn1]{Current address: Albert Einstein Center for Fundamental Physics, University of Bern, 3012 Bern, Switzerland}

\address[qu]{Department of Physics, Engineering Physics \& Astronomy, Queen's University, Kingston, Ontario, Canada, K7L 3N6}

\address[tum]{Physik-Department E15, Technische Universit\"{a}t M\"{u}nchen, Garching, Germany 85748}

\address[pi]{Perimeter Institute for Theoretical Physics, 31 Caroline St. N, Waterloo, Ontario, Canada N2L 2Y5}

\address[mac]{Department of Physics \& Astronomy, McMaster University, 1280 Main St. W, Hamilton, Ontario, Canada, L8S 4M1}

\begin{abstract}
Scintillating calorimeters are cryogenic detectors combining a measurement of scintillation with one of phonons to provide particle identification. In view of developing alkali halide devices of this type able to check the DAMA/LIBRA claim for the observation of dark matter, we have simulated detector performances to determine their sensitivity  by two methods with little model-dependence. We conclude that if performance of the phonon channel can be brought in line with those of other materials, an exposure of 10~kg-days would suffice to check the DAMA/LIBRA claim in standard astrophysical scenarios. Additionally, a fairly modest array of 5~kg with background rejection would be able to directly check the DAMA/LIBRA modulation result in 2~years.
\end{abstract}

\end{frontmatter}

%\linenumbers

\section{Introduction}
The mystery of dark matter has been open since 1933~\cite{zwicky1933rotverschiebung}: most of the matter in the Universe only appears through its gravitational interactions~\cite{PhysRevD.86.010001}.  New particles predicted beyond the standard model of particle physics may provide a solution.  Detection of such particles is a challenge because of the low energies and interaction cross-sections involved~\cite{schnee_introduction_2011}.
A convincing claim of dark matter detection has yet to be realized, but several experiments~\cite{Agnese:2013x,PhysRevD.88.012002,Angloher:2012tx}, including DAMA/LIBRA~\cite{Bernabei:2008ec,Bernabei:2013ax} (referred to as DAMA henceforth), have observed event excesses inconsistent to various degrees with known backgrounds, in tension with other experiments~\cite{PhysRevD.86.010001}. The complete DAMA experiment consists of 25 NaI(Tl) detectors totaling 250~kg and  operating at room temperature in a low-background environment.  The signal read from each crystal is the scintillation light created by particle interactions, for instance the nuclear recoil from the elastic scattering of a dark matter particle.  The experiment has been designed to identify dark matter via the annual modulation signature caused by yearly variations in the speed of the Earth with respect to our dark matter halo~\cite{Bernabei:2008ec,Bernabei:2013ax}. DAMA observes a statistically-robust modulation consistent with an astrophysical origin and that they claim is unexplainable as background, though the required modulation fraction appears to be very large~\cite{pradler_unverified_2013,Pradler2013168,Kudryavtsev:2010bj}. 
Moreover, under standard astrophysical and particle assumptions, this claim is incompatible with other direct detection experiments employing different techniques and targets, such as XENON100~\cite{Aprile:2012x}, CDMS-II~\cite{Ahmed:2010sc}, SuperCDMS~\cite{Agnese:2014lm}, CDMSlite~\cite{Agnese:2014prl}, EDELWEISS~\cite{Armengaud:2012x}, LUX~\cite{Akerib:2014jv} and PICASSO~\cite{Archambault:2012lm}. 

In comparison to the simple scintillation detectors used by DAMA, which do not provide event-by-event background discrimination, scintillating calorimeters can give more insight into the nature of the interacting particle. A scintillating calorimeter is a particle detector consisting of a scintillating crystal held at cryogenic temperatures ($\lesssim$~50~mK) as the target medium, read-out by a light detector and thermal sensors, such that, for a given particle interaction, both scintillation and phonon signals can be observed~\cite{schnee_introduction_2011}. Nuclear recoils will produce less scintillation light for a given energy deposit than ionizing radiation in the form of alphas, gammas, and betas (through a process known as quenching), which allows for very powerful discrimination against background events. The CRESST experiment uses an array of scintillating calorimeters to detect nuclear recoils from dark matter particles~\cite{Angloher:2012tx}.  
A scintillating calorimeter capable of background discrimination and based on NaI could shed light on the controversial DAMA claim by potentially revealing target-specific backgrounds or interactions.  This approach would be complementary to the various proposals and attempts to test DAMA using scintillation-only NaI, for instance DM-Ice, which is located at the South Pole to study the phase of a possible signal~\cite{Cherwinka:2014xta},  the SABRE project, located in a background-rejecting veto~\cite{Xu:2014}, ANAIS~\cite{amare_preliminary_2014} and KIMS~\cite{Kim:2014x}.  Moreover, irrespective of DAMA, alkali-halide scintillating calorimeters could help explore new WIMP parameter space~\cite{cerdeno_scintillating_2014}.

In this paper, we  simulate the expected background discrimination power and sensitivity to WIMPs of a scintillating calorimeter with an alkali halide target for different exposures and under various assumptions about phonon detector resolutions. We also study the sensitivity of a background-rejecting detector such as this to an annual modulation signal.  Lastly, the technical feasibility of this type of detector is explored.

\section{Expected performance of alkali halide scintillating calorimeters}
\label{sec:detector_performances}

In this section, we discuss what performances can be expected from alkali halide detectors in terms of scintillation and phonon signals.
We base our design on CRESST scintillation-phonon devices~\cite{Angloher:2012tx}. The standard cylindrical size of the scintillators (height and diameter 4~cm) yields an individual detector mass of 184~g for NaI (density 3.67~g/cm${}^3$) and 227~g for CsI (density 4.51~g/cm${}^3$).  An array  of several of these could be assembled, though it would be worthwhile to develop larger, kg-scale (height and diameter 7~cm for NaI), individual devices, that already exist as scintillators.  
For the light channel, we assume standard CRESST performances, including a light collection efficiency (fraction of photons created by the scintillation event that are detected) of 31\%~\cite{Kiefer:2012va}, and an energy-independent contribution from the light detector to the resolution with a standard deviation of 10~eV~\cite{Schmaler:2010up,Kiefer:2012va,Roth:2013uz}.
The scintillation of NaI~\cite{VanSciver:1958,Cooke:1964da,West:1970wq,Spooner:1996cf,Moszynski:2003ie,Sibczynski:2010df,sibczynski_properties_2011} and CsI~\cite{Kubota:1988,Schotanus:1990tk,Amsler:2002kq,Moszynski:2005hu,vm_CsI} (with and without doping) have been studied over a range of temperatures. The light yield (the amount of light emitted by the scintillator for a given deposited energy) 
tends to increase as the temperature of the scintillator decreases. 
More recent work has gone into studying the light yield of NaI and NaI(Tl) under alphas down to 1~K~\cite{sailer2012low}, and studying the optical properties of NaI in the 1~K range~\cite{coron2013study}. The light yields we use in the following come from our own study of the alpha and gamma scintillation of these crystals~\cite{Nadeau:2014phd}. More information on our experimental setup can be found in~\cite{verdier_2.8_2009,Verdier:2011pb,di_stefano_counting_2013}. The results for gamma scintillation at our lowest temperature, 3.4~K, are shown in Table~\ref{tab:absLY}. 
The light yield seems to vary little around this temperature, hence our using these results in the tens of millikelvin range relevant for cryogenic detectors. We note that the only sample for which cryogenic data exist is CsI. Our experimental value of its light yield, when combined with the light detection efficiency and the photon energy described below in Sec.~\ref{sec:DiscrimBands},
leads to a fraction of deposited energy detected as light of $59 \mbox{ ph/keV} \ \times \ 3.9 \mbox{ eV/ph} \  \times \ 0.3 \approx 0.07$. This is consistent with the 7.1\% of energy directly measured as light in cryogenic detectors~\cite{Schaffner:2012ei}.

%Table 1, Single Column
\begin{table}
\centering
\caption{\label{tab:absLY}Light yield from 60~keV $\gamma$ interactions for the alkali halides at 3.4~K, based on measured spectra~\cite{Nadeau:2014phd}.  All values are expressed in photons per keV.}
\bgroup
\def\arraystretch{1.5}
\begin{tabular}{@{}rccc@{}}
\hline\hline
 & NaI & CsI & NaI(Tl)\\
\hline
LY & $19.5 \pm 1.0$ & $58.9 \pm 5.6$ & $40.6 \pm 0.8$\\
\hline\hline
\end{tabular}
\egroup
\end{table}

The performance of the phonon channel is expected to depend strongly on the absorber.  
Also, as direct deposition of the tungsten phonon sensors on the alkali halides may be problematic because of the low melting temperature of these materials, these detectors can benefit from the CRESST composite technology in which the sensor is deposited on a small substrate then glued on the main scintillator~\cite{kiefer_comp}.
The only published alkali halide results we are aware of are for CsI, and suffer from very poor phonon resolution of 11.6~keV for test pulses corresponding to an energy of 139~keV~\cite{Schaffner:2012ei}.  This is taken as the energy-independent, dominant, baseline noise.
Possible explanations for this include the low Debye temperature of CsI (128~K~\cite{Marshall:1969gl}), or surface degradation due to hygroscopicity~\cite{Schaffner:2012ei}.  Possibilities to remediate this have not been exhausted and include optimization and adjustment of the utilized temperature sensor (concerning its size, the material and gluing technique used to attach it to the crystal).  
Therefore in the rest of this analysis, we consider two scenarios for detector performance.  
In the first, optimistic, case, we assume that performances can be brought into line with those from current CRESST CaWO$_4$ detectors with glued sensors~\cite{Lang:2010bt}.  We interpolate their phonon resolution by a function that depends on the square root of the energy:
\begin{equation}
\label{eq:phonRes_1}
\sigma_P (E) = \sigma_P(0) + \sqrt{\frac{E}{122 \mbox{ keV}}} (\sigma_P(122\mbox{ keV}) - \sigma_P(0)).
\end{equation} 
with  a baseline noise of $\sigma_P(0)=210$~eV,  and a resolution at high energies (122~keV) of $\sigma_P$(122 keV)~$=840$~eV. 
The system is assumed to be triggered by the phonon channel with a threshold given by 5 times the standard deviation of the baseline noise.
In the second case, we consider the resolution to be constant, with  $\sigma_P = 10$~keV, close to the current value~\cite{Schaffner:2012ei}.  This corresponds to the small pulse size making the baseline resolution the dominant term.
In this case, we assume that the the good performances of the light channel enable the phonon threshold to be kept at 2 times the standard deviation of its baseline noise.

\section{Time-independent sensitivity limited by background leakage into signal region}

Our first approach involves considering a canonical WIMP and astrophysical scenario, and applying a standard analysis based on time-independent rates~\cite{schnee_introduction_2011}:
i) assuming that all of the background is time-independent and caused by electron recoils, ii) estimating how much of the background leaks into the nuclear recoil region of our detectors and iii) determining how the leakage would limit the sensitivity of these detectors to nuclear recoils from a WIMP.
Since, for NaI and NaI(Tl), we are considering the same target as DAMA, this method is less sensitive to assumptions on particle couplings, which are being scoured for loopholes to reconcile experimental results coming from DAMA on the one hand, and  detectors with other target materials on the other~\cite{Smith:2001dm,Chang:2010,Buckley:2013jd,Foot:2013by,Gresham:2014cd}.

\subsection{Background models}
\label{sec:BGModels}
We use the reported background spectra from the DAMA and KIMS (2005) experiments as the background models for our Monte Carlo simulations. The DAMA background spectrum (used for the simulations with NaI(Tl) and NaI) is for an exposure of 0.53 ton-yr, taken from~\cite{Bernabei:2008ec} and studied in~\cite{pradler_unverified_2013,Pradler2013168,Kudryavtsev:2010bj}. The spectrum has a flat component of roughly 1 cpd/kg/keV with a peak at 3.2~keV due to $^{40}$K decays. The background spectrum for KIMS (an array of CsI(Tl) detectors) for an exposure of 237 kg-d is taken from~\cite{Lee:2005ax} and is used for our CsI simulations. The KIMS background has a flat component of roughly 7 cpd/kg/keV. 
Both background spectra have an exponential component at low energies, which we have included in this analysis, though they are most likely due to PMT noise. We also force the spectra to remain flat out to 60 keV, though this has little impact on low-mass WIMPs and is not quite the case for DAMA~\cite{Bernabei:2009ec,pradler_unverified_2013}.

\subsection{Generating discrimination bands}
\label{sec:DiscrimBands}
%
%Figure 1, Single Column
\begin{figure}[h]
		\centering
		\includegraphics[width=0.8\columnwidth]{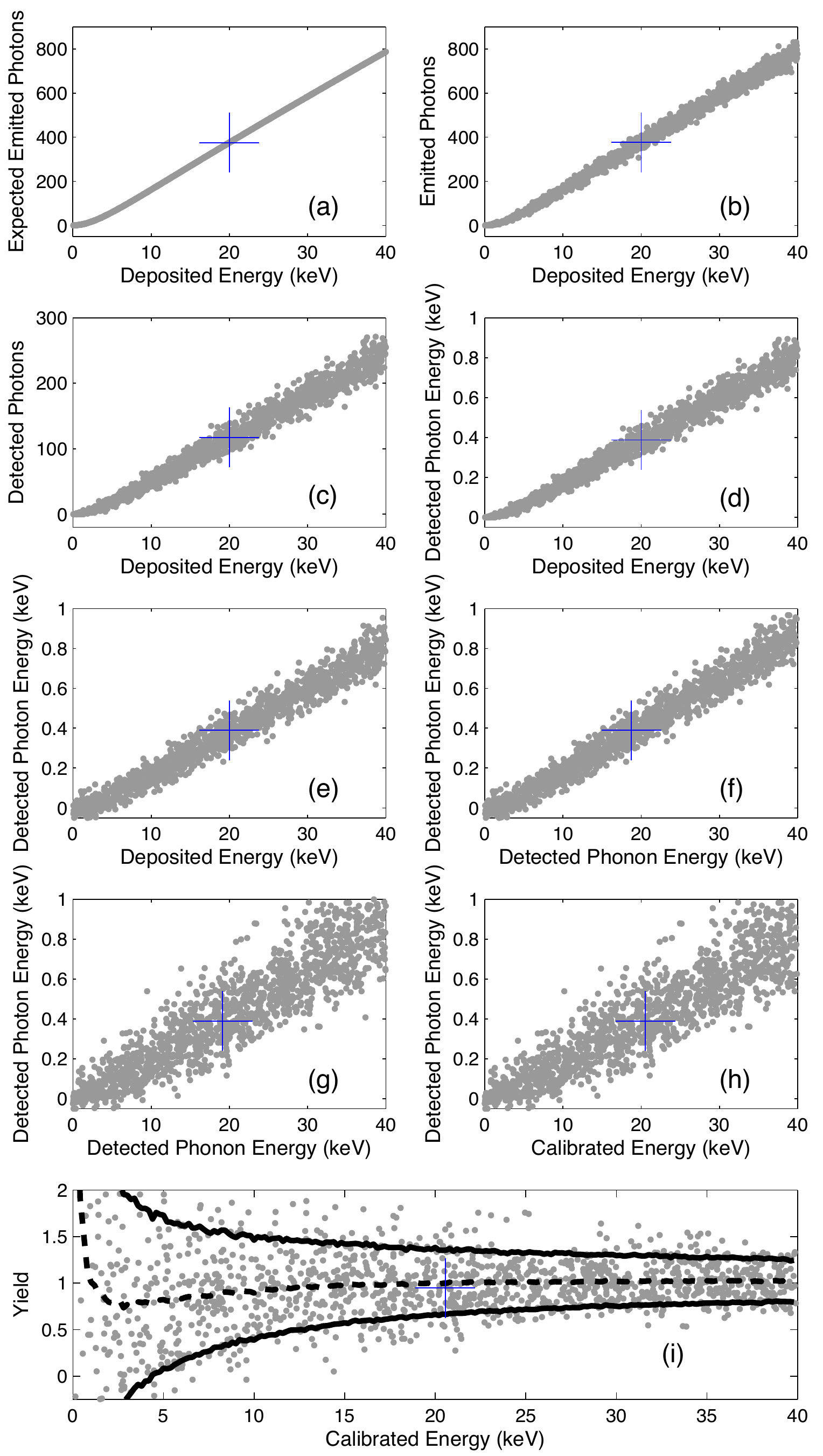}
		\caption{\label{fig:testPlots} Generation of discrimination bands, illustrated on NaI for the electron/$\gamma$ band. See Sec.~\ref{sec:DiscrimBands} for details of procedure. For illustrative purposes, $\sigma_L =30$~eV and the model in Eq.~\ref{eq:phonRes_1} has been used here with $\sigma_P$(0~keV)~$=500$~eV and $\sigma_P$(122~keV)~$=10$~keV.  Blue cross demonstrates the procedure for the average of a group of 20~keV events.}
\end{figure}
Background discrimination is carried out in a normalized light yield vs energy plane.  The regions covering a given fraction of the signal and those covering a given fraction of the background are referred to as discrimination bands.
We generated these bands
for the interaction of $\gamma$ quanta and the respective nuclear recoils possible in each considered crystal using Monte Carlo simulations.  For each material, the inputs are the LY from Table~\ref{tab:absLY}, the nuclear quenching factors $Q$ (assumed here to be energy-independent), the mean energy of the emitted photons ${\cal E}$, the light collection efficiency $\epsilon$ of the detector module, and resolutions of the light detector and phonon detector, $\sigma_L$ and $\sigma_P$, respectively. The algorithm is illustrated step-by-step in Figure~\ref{fig:testPlots} and described in the following list for the example of electron recoils ($Q:=1$), where each item of the list corresponds with the appropriate subfigure:
\begin{enumerate}[a)]
\item Randomly draw $10^7$ particle event energy deposits $E$ from a uniform distribution between 0 and $E_{max}$.   For each of these deposits, calculate the expected number of photons $\nu$ emitted by the scintillator: 
\begin{equation}
\label{eq:N}
\nu = E \times LY \times Q \times \eta(E). 
\end{equation} 
The energy-dependent factor $\eta(E)$ accounts for the nonlinearity of the scintillator. It has been calculated using a model proposed in~\cite{Payne:2009en} for the nonlinear response of inorganic scintillators to electron interactions and fit to measurements of the nonlinearity of NaI(Tl)~\cite{Payne:2011ie}, NaI~\cite{Moszynski:2003ie} and CsI~\cite{Moszynski:2005hu}. 
\item For each of the previous events, generate the actual number of emitted photons, $n$, from a Poisson distribution with expectation $\nu$.
\item For each event, determine the number of detected photons from a binomial distribution with $n$ attempts and a success probability equal to the light collection efficiency $\epsilon$. 
\item Convert to detected photon energy by multiplying by the mean energy ${\cal E}$ of the emitted photons for each crystal (NaI: 3.3~eV~\cite{Cooke:1964da}, NaI(Tl): 2.95~eV~\cite{Cooke:1964da}, CsI: 3.9~eV~\cite{Schotanus:1990tk}).
More recent work on NaI(Tl)~\cite{coron2013study} identifies two peaks at 2.82~eV and 3.76~eV at 4~K; however, in our case, the actual photon energy does not have much effect on the  width of the band which is dominated by  phonon channel resolution  (see g) below).
\item Account for the light detector resolution $\sigma_L$ by applying a random fluctuation from a normal distribution with standard deviation $\sigma = \sigma_L$ to the detected light energy $E_L$. 
\item Convert the deposited energy $E$ to detected phonon energy by subtracting the emitted light energy: $E_P = E - n {\cal E}$.
\item Apply the phonon detector resolution $\sigma_P$ to the detected phonon energy 
as a random fluctuation from a normal distribution with $\sigma = \sigma_P$. This is the quantity that would be measured in the phonon channel of an actual experiment.
\item Calibrate the phonon energy scale using the expected value at 122~keV, as would be performed in an actual experiment.  
\item Convert to a yield vs. calibrated phonon energy plane, where yield $Y$ is defined as 
$Y = E_L/E_P$ (both including resolutions), and determine the discrimination band for electron and $\gamma$ events by binning with a width of 200~eV along the phonon energy axis, and calculating the 10th and 90th percentiles along the yield axis within each bin. The yield is normalized to its value for 122~keV $\gamma$ quanta.
\end{enumerate}

Discrimination bands for the different nuclei in each target crystal are determined in the same manner as above, except the nuclear quenching factor $Q$ 
for each nucleus is applied when calculating $\nu$ in Eq.~\ref{eq:N},
resulting in the discrimination bands shown in Figures~\ref{fig:discrimBands} (optimistic phonon resolution) and \ref{fig:discrimBands3} (current phonon resolution). 
Though there remains some disagreement between various quenching factor measurements at low energies for NaI(Tl) (c.f.  Fig.~9 in~\cite{Collar:2013vd}), we have taken the following values for the nuclear recoil quenching factors, neglecting energy and temperature dependence: 0.3 for Na and 0.1 for I in NaI(Tl)~\cite{Bernabei:1996fg,Gerbier:1999ap}.  We carry over the same values to NaI since we are unaware of any measurements on that undoped material.
Similarly, we take values from CsI(Tl) for CsI: 0.1 for Cs and I in CsI~\cite{Wang:2002bv}. 
Lastly, we neglect any possible effects of gamma quenching at low energies, which exists in CaWO$_4$ and results in a reduced light yield for $\gamma$ quanta compared to equally energetic electron recoils~\cite{Lang:2009ua}.

\subsection{Expected sensitivities}
\begin{figure}
		\centering
		\includegraphics[width=\columnwidth]{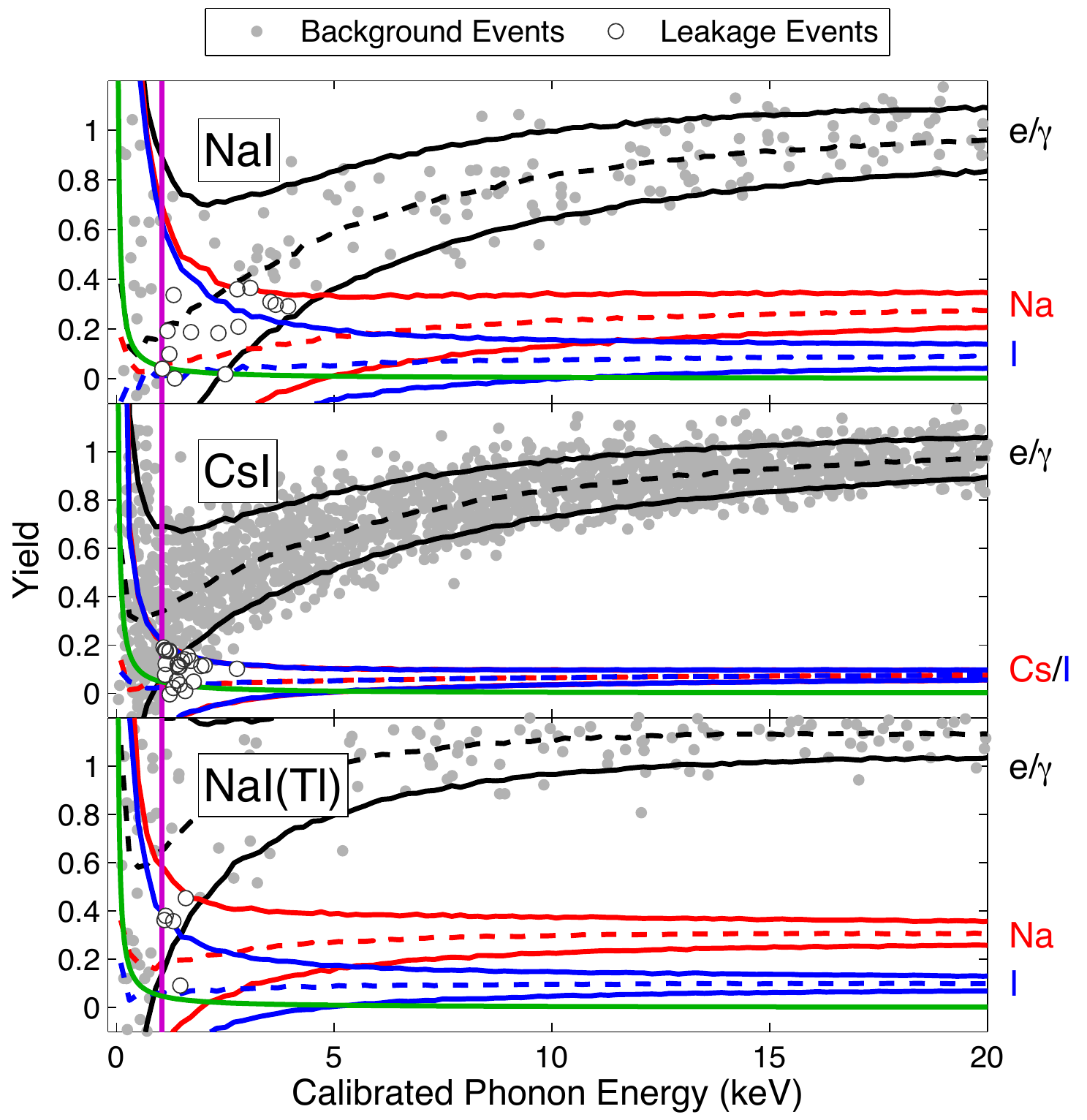}
		\caption{\label{fig:discrimBands} Discrimination bands generated for NaI, CsI and NaI(Tl) with light detector resolution of 10~eV and optimistic phonon detector resolution 210~eV at 0~keV and 840~eV at 122~keV. Events are simulated from the DAMA (NaI) and KIMS (CsI) background distributions for 10~kg-days between 0-20~keV with a phonon energy threshold of 1.05~keV (vertical, magenta line), which is 5 standard deviations of the baseline noise.   Five standard deviations of the light detector baseline noise are represented as the green curve.}
\end{figure}

\begin{figure}
		\centering
		\includegraphics[width=\columnwidth]{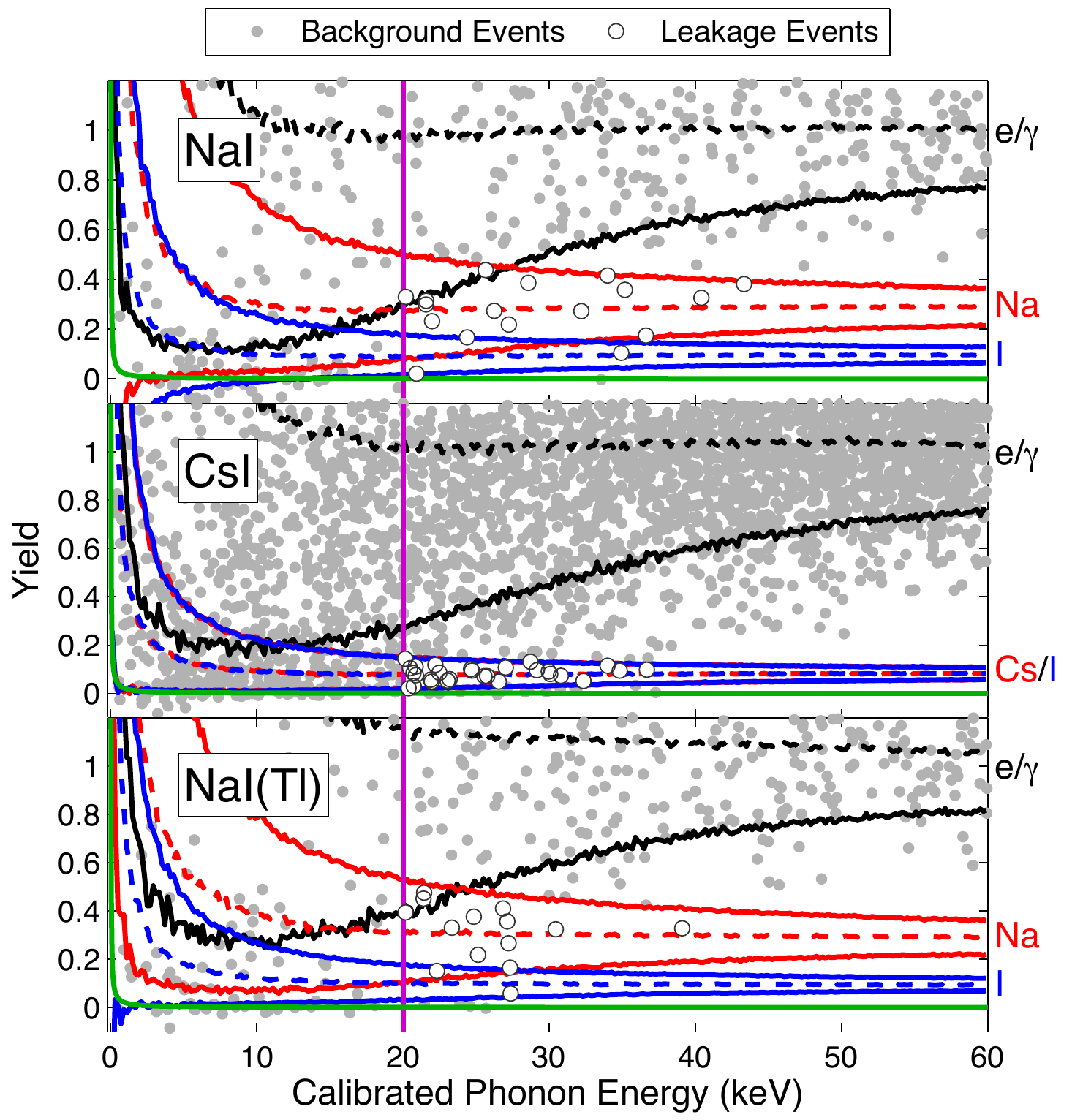}
		\caption{\label{fig:discrimBands3} Same as Figure~\ref{fig:discrimBands}, but with  current phonon detector resolution of 10~keV, with bands calculated up to 60~keV, and with phonon energy threshold of 20~keV (vertical, magenta line), which is 2 standard deviations of the baseline noise.    
        Signal region is also at least five standard deviations above the light detector baseline noise (green curve).}
\end{figure}

For each target material and exposure, a set of energy deposits are randomly drawn from the background distributions (assumed to be all electron and $\gamma$ events) described in Section~\ref{sec:BGModels}, and then converted to the Yield vs.\ Deposited Energy parameter space by the procedure described in the previous section. 
As  illustrated in Figures~\ref{fig:discrimBands} and \ref{fig:discrimBands3},
the bands and background events 
are compared to each other to determine how many electron/$\gamma$ background events leak into the nuclear recoil bands, where we would expect WIMP events to occur.
The energies of all of these events are saved.  However, in this analysis, no attempt is made to use the yield values for a more sophisticated background identification. In the case of the current phonon resolution (Fig.~\ref{fig:discrimBands3}), the signal regions are above a virtual light detector threshold that is 5 standard deviations of its baseline noise.
%

%Figure 2, Single Column

%Figure 3, Single Column

With our collection of leakage events determined, we predict the sensitivity to WIMPs that our proposed detector may have for given exposures and detector resolutions using the high statistics extension to the Optimum Interval Method~\cite{Yellin:2002pd,Yellin:2007ax}. The method computes the 90\% confidence level (CL) upper limit of a signal for an unknown background. These calculations are performed under typical astrophysics assumptions (WIMP mass density of 0.3~GeV/c$^2$/cm$^3$, mean WIMP velocity with respect to the galaxy of 220~km/s, mean circular velocity of Earth with respect to the galactic center of 232~km/s, and galactic escape velocity of 544~km/s~\cite{Smith:2006ym}) and over a range of WIMP masses to determine the expected WIMP-nucleon cross-section for each WIMP mass. 

\begin{figure}
		\centering
        \includegraphics[width=\columnwidth]{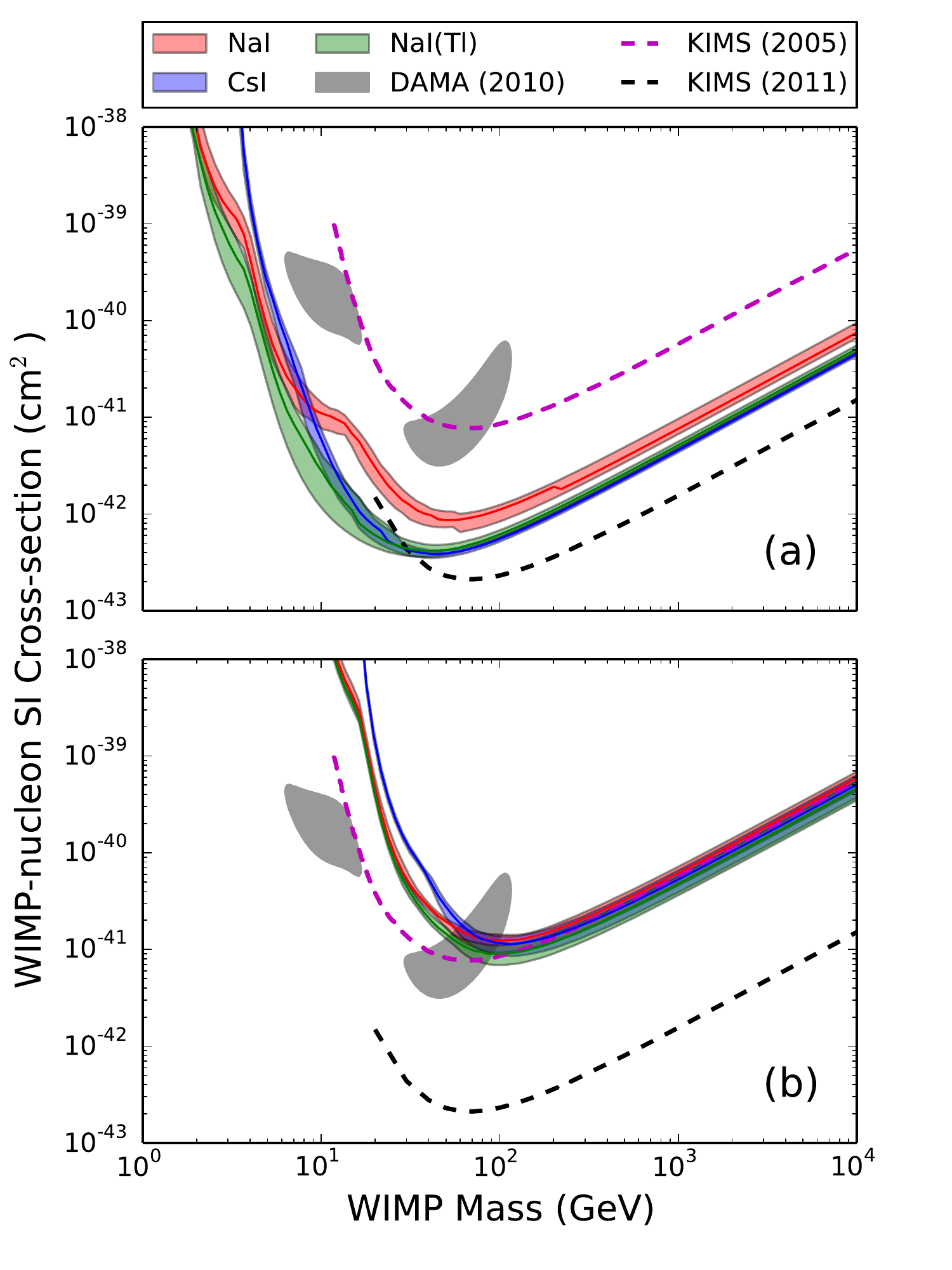}
		\caption{\label{fig:expLim1} Expected sensitivities for different alkali halide calorimeters, assuming DAMA and KIMS 2005 backgrounds, with light detector resolution of 10~eV and (a) optimistic phonon performances (840~eV at 122~keV) and a 10~kg-days exposure, and (b) current phonon performances (baseline noise dominated resolution of 10~keV) and 100~kg-days. 
        Also shown are the 2005 (237~kg-days)~\cite{Lee:2005ax} and 2011 (24~tonne-days)~\cite{Kim:2012rza} KIMS limits.
        The optimistic scenario allows investigation of the DAMA claim~\cite{Savage:2009hi} with a small exposure.  In the case of current performances, sensitivity is background-limited before the full DAMA region can be explored.  Filled-in, translucent regions cover the 10th and 90th percentiles and  the median of a batch of simulated experiments generated according to our model.}                
\end{figure}

To determine how well this proposed detector could test the DAMA dark matter detection claim, we compare the expected sensitivity curves for each of our alkali halide targets with the DAMA discovery contours for Na and I~\cite{Savage:2009hi}, shown in Figure~\ref{fig:expLim1}.
We also show limits from the KIMS experiment~\cite{Kim:2012rza} as other comparisons with our results (the 2005 limit with the background we have used, and the more recent and improved 2011 limit).
%
%Figure 4, Single Column
Figure~\ref{fig:expLim1}(a) is the result of our simulations for an alkali halide detector after 10~kg-days of exposure.
For the optimistic phonon detector performance scenario described in Sec.~\ref{sec:detector_performances}, 
this modest exposure would  check the entire DAMA claim.  For the current performance scenario ($\sigma_P$(0~keV) $ = \sigma_P$(122~keV) $ = 10$~keV), background discrimination is degraded as seen in Fig.~\ref{fig:discrimBands3}. The sensitivity is background-limited after 100~kg-days of exposure, illustrated in Fig.~\ref{fig:expLim1}(b).
More sophisticated analysis techniques, such as likelihoods, may be able to better exploit the yield information to distinguish different types of recoils and produce more stringent limits.
Assuming current phonon performances and a background lowered to the level of that expected by the SABRE collaboration~\cite{Xu:2014}, an exposure of 100~kgd would allow to check the iodine-recoil, higher-mass DAMA contour.

Regarding the relative sensitivities of pure and doped NaI, the latter does marginally better thanks to its light yield which is roughly twice that of the former at low temperatures, enabling slightly better background rejection.
CsI is less competitive than both varieties of NaI at low WIMP masses since its heavier nuclei  are a worse match from the standpoint of scattering kinematics.  
For heavy WIMPs, CsI is comparable to NaI(Tl) because its higher background is offset by better background rejection stemming from the smaller quenching factors of Cs and I compared to Na  (Figures~\ref{fig:discrimBands} and~\ref{fig:discrimBands3}).

\section{Sensitivity to a generic modulation}
Our second approach is even less model-dependent, as it only involves looking for a modulation in time similar to that of DAMA.
The DAMA claim is based on the statistically-robust detection of an annual modulation of the detected event rate interpreted as dark matter.   In the 2--6~keV$_{\mathrm{ee}}$ (electron equivalent) energy range, the observed modulation amplitude is $A=0.0448$~cpd/kg, the phase is $t_0=144$~days after January~1, and the period is $T=365$~days~\cite{Bernabei:2013ax}.  The exposure is very large, of the order of 1~tonne-yr. In this section, we study what sensitivity smaller, background-rejecting, detectors with low enough threshold would have to this modulation. Since the DAMA experiment has no background rejection, though the total combined signal and background DC level is known ($\approx 1$~cpd/kg/keV), the DC level of the signal itself can not be known without resorting to astrophysical and particle models.
Several different DC levels are simulated to cover various possible combinations of DC signal and DC background leakage.  We assume that any modulation is a result of the dark matter signal and not the leaking background.  
For the known AC signal level, we choose a DC signal+background level, and for a given detector mass and exposure, generate a group of Monte-Carlo (MC) simulated datasets using the above values of $A$, $t_0$, and $T$.   For each MC, the number of events $n$ and their arrival dates $t_i$ are drawn from the  DC+AC expectation. These MCs are then analyzed by two methods that try to detect the modulation.  

\subsection{Likelihood ratio test}
\label{sec_LLR}
The first method we use is a likelihood ratio test.
We define a modulation function $f(t;\theta)$ as a function of time $t$ and modulation parameters $\theta$:
\begin{equation}
f(t;DC,\alpha,t_0) = DC \left\{ 1 + \alpha \sin \left(2 \pi \frac{ t - t_0}{T} \right) \right\}.
\label{Mike_modeq}
\end{equation}
where $DC$ is the time-independent signal+background contribution, and $\alpha = A/DC \geq 0$ is the ratio of AC modulation amplitude to DC level.
Using this function, we define an unbinned extended likelihood function~\cite{cowan_statistical_1998}:
\begin{equation}
L(\theta) = \frac{\mu (\theta)^n}{n!}e^{-\mu (\theta)}\prod\limits_{i=1}^{n} g(t_i;\theta)
\label{Mike_Leq_1}
\end{equation}
where $n$ is the total number of data points, and  $g \equiv f/\mu$ is the probability density function of $f(t;\theta)$ with an expected number of events $\mu$ defined as 
\begin{equation}
\mu(DC,\alpha,t_0) = DC \left\{ \tau + \frac{T}{2\pi } \alpha \left[ \cos \left( 2 \pi \frac{ t_0}{T} \right) -  \cos \left( 2 \pi \frac{ \tau - t_0}{T} \right) \right] \right\}
\end{equation}
i.e. the total integral of $f(t;\theta)$ over the exposure time $\tau$. 

For each data MC, the likelihood function is maximized for the modulation hypothesis, with the DC level, phase and modulation amplitude as free parameters, but the period fixed to one year, leading to a value of $\hat{L}_1$. It is separately maximized for the null hypothesis over the DC level only, with the modulation amplitude set to 0, leading to a value of $\hat{L}_0$. 
The ratio of likelihoods $\frac{\hat{L}_1}{\hat{L}_0}$ tells us how sure we are that a claimed modulation is due to a true modulation and not a statistical fluctuation of a null-hypothesis background. We have verified on another set of 5000 MCs with no modulation that the distribution of $ 2 \ln (\frac{\hat{L}_1}{\hat{L}_0})$ follows a $\chi^2$ distribution with 2 degrees of freedom, as expected for sufficient statistics~\cite{Wilks:1938}.
This enables us to  quantify the result in terms of p-value (the chance that a statistical fluctuation of the null hypothesis generates a greater likelihood ratio than the actual data MC).  The detection certainty, or confidence level for a detection, would be $1-p$.

\begin{figure}
   \begin{center}
   \includegraphics[width=0.8\columnwidth]{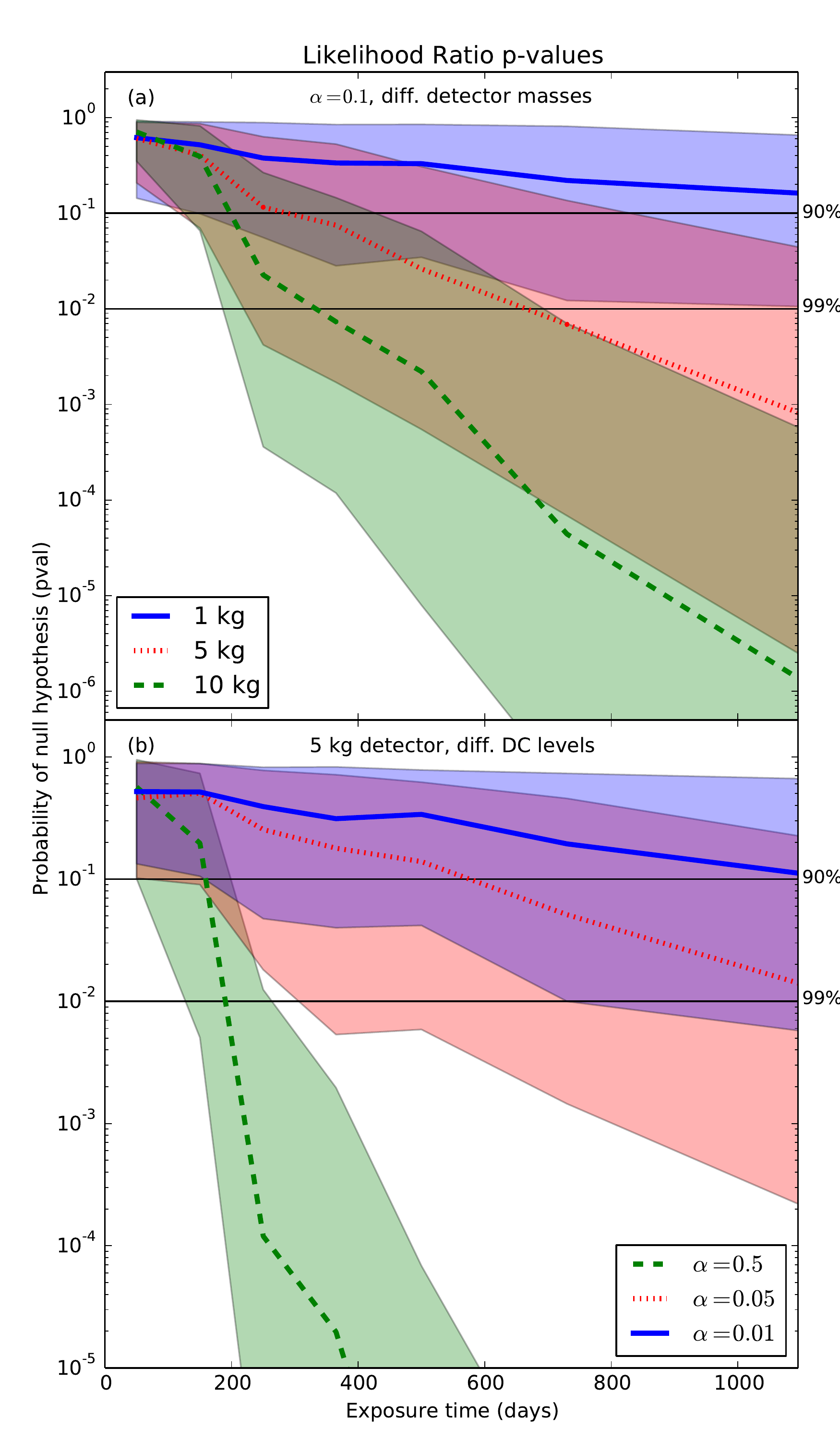}\\
   \end{center}
    \caption{(a) p-Values calculated using the likelihood ratio method, for various masses of detector, as a function of exposure time.  The DC level has been set to 10 times the modulation amplitude.  (b) p-Values calculated for a 5~kg detector but adjusting the DC background level added to the DAMA modulation, effectively changing the value of $\alpha$.  With the same background level as DAMA and a purely AC signal, $\alpha=0.5$ corresponds to a 99\% background rejection, whereas $\alpha=0.01$ would correspond to no background rejection at all. For the $\alpha=0.5$ calculation, the line drops off the plot as the confidence level quickly approaches 100\%.} 
    \label{Mike_pVals}
\end{figure}

For each exposure scenario, 500 data MCs with a modulation were generated and subject to the likelihood ratio procedure.
The median and 10th and 90th percentiles of the p-value were recorded for several exposure times and detector masses, and are displayed in Figure~\ref{Mike_pVals}.
Fig.~\ref{Mike_pVals}(a) illustrates p-values plotted as a function of exposure duration, for various detector masses, and a fixed value of $\alpha = 1/10$.
%
%Figure 5, Single Column
This shows that a 5~kg detector, or array of detectors, has a better than 50\% chance of observing such a modulation in roughly 2~years at a 99\%~CL.  As expected, a more massive array of detectors (10~kg) would have a stronger chance ($\approx90\%$) of confirming the modulation at the same CL in the same time.  Conversely, a 1~kg array would struggle to obtain a 90\%~CL discovery even in a three-year run.

Results for a 5~kg detector with various DC levels as a function of exposure time are shown in Fig.~\ref{Mike_pVals}(b).  If we assume a minimal DC signal contribution equal to the AC signal contribution, a total DC to AC ratio of 2 $(\alpha=0.5)$ corresponds to at least 99\% background rejection, when compared with DAMA's 
time-independent level of 1~cpd/kg/keV.
In this case, the modulation has a better than 90\% chance of being detected in less than 1 year, with better than a 99\%~CL.
Other ratios are also shown in this figure, corresponding to 80\% background rejection for $\alpha=0.05$ and little, if any, background rejection for $\alpha=0.01$, again assuming a minimal DC signal.
In the latter case, the detector would have a 50\% chance of detecting a signal at 90\% certainty after 3~years.
This case with little or no background rejection is  also relevant for an electron-recoil modulation caused by some background or exotic type of dark matter, assuming there is no nuclear-recoil type background, or, more generally, for any type of non-discriminating detector trying to check DAMA starting from the same background level.
Lastly, a lower DC level and/or a larger detector will increase the CL at a faster rate, as  expected.

\subsection{Lomb-Scargle analysis}
We next apply the Lomb-Scargle method~\cite{scargle_studies_1982} to our MCs.  This method, like the standard Fourier one, can be used to obtain a power spectrum of a data set as a function of a time scale (periodogram); however, it has the advantage that it can be applied to unevenly sampled data.  
In addition, the method allows to quantify the probability that a given feature in the periodogram is a fluctuation of the null hypothesis, as opposed to a true periodic feature.  Since in our case we are looking for a modulation with a known period of one year, the null hypothesis can be rejected with a confidence level $1-p_0$ provided the power at that frequency is at least $- \ln p_0$~\cite{scargle_studies_1982} (this simple criterion would not hold were we looking for power at an unknown frequency). 

\begin{figure}[h]
   \begin{center}
   \begin{tabular}{c}
   \includegraphics[width=0.9\columnwidth]{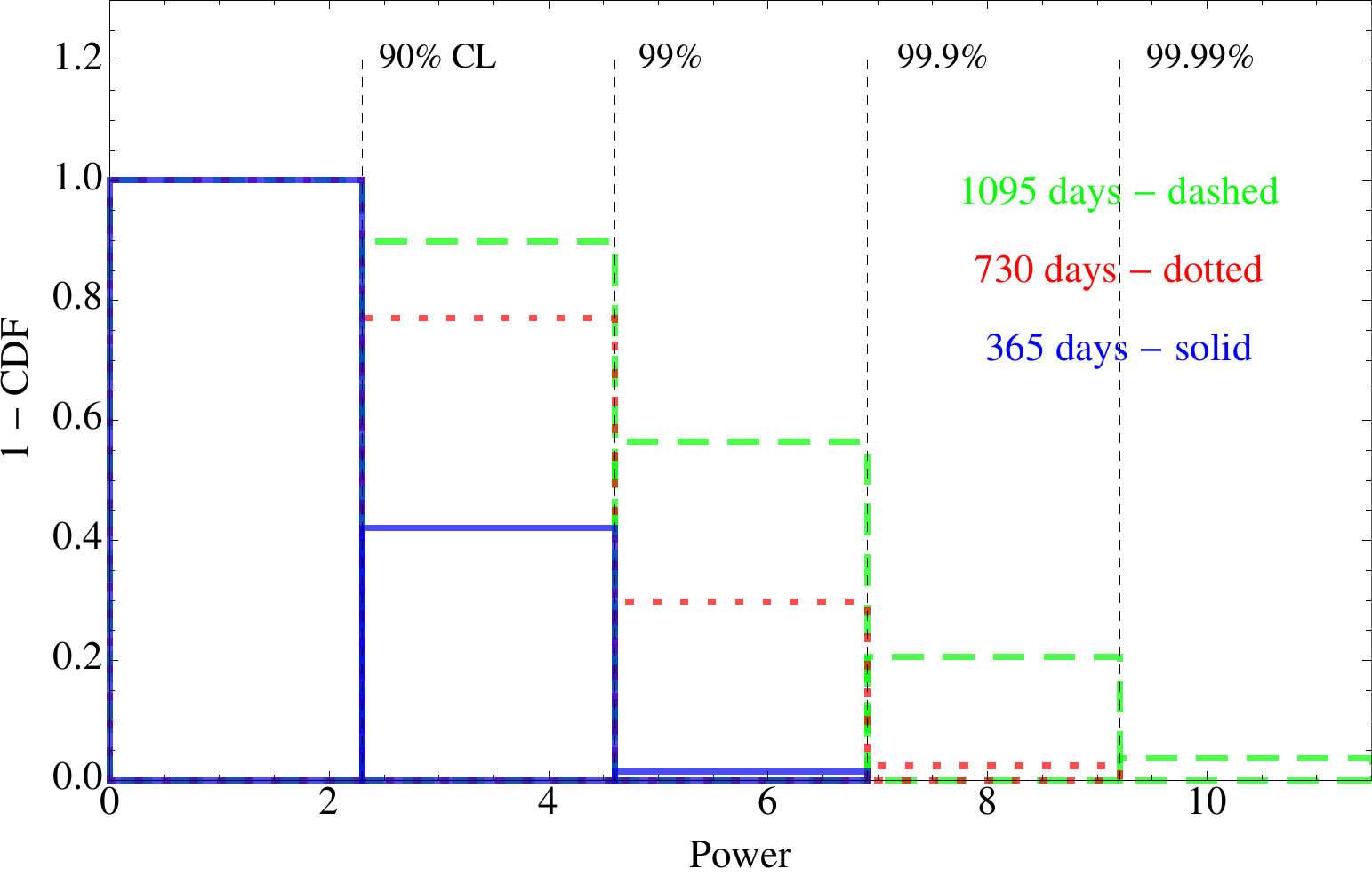}
   \end{tabular}
   \end{center}
  \caption{Complement of cumulative distribution of the Lomb-Scargle analysis applied to simulations of different exposure durations for a 5~kg detector with a DC level equal to ten times the AC one.  Power thresholds corresponding to a particular confidence level appear as vertical lines. The analysis shows that for a 2-year exposure, there would be roughly a 75\% chance 
  of seeing a modulation at a 90\%~CL.  For a 3-year exposure, there would be about a 55\% chance of seeing a modulation at 99\%~CL.
  } 
    \label{fig:LS_1}
\end{figure}

We have evaluated the expected sensitivity of the Lomb-Scargle test statistic by applying it to our data MCs, generated with a one-year modulation,  to obtain the expected distribution of power with a period of one year. 
The results are shown in Fig.~\ref{fig:LS_1} where we plot a complement of the cumulative distribution of the number of MCs yielding power in a certain interval. 
The necessary power thresholds corresponding to rejection at a certain CL are shown as vertical lines.  
%
%Figure 6, Single Column
We note that the confidence level derived from the data using the Lomb-Scargle method cannot be compared directly to the likelihood ratio discussed in the previous section since the two methods ask different statistical questions. In particular, the Lomb-Scargle periodogram does not require an assumption about the form of the putative periodic signal, unlike our likelihood ratio which assumes the modulation to be sinusoidal. Therefore, it is generally less powerful than the likelihood ratio when such an assumption is made.
This is evident for instance in the 5~kg, 2~year exposure with an AC-to-DC ratio of $\alpha=0.1$.  The likelihood ratio (Fig.~\ref{Mike_pVals}(a))  provides a $\approx 50\%$ chance of seeing a sinusoidal modulation at better than 99\%~CL, whereas the Lomb-Scargle method gives roughly only a 30\% chance of detecting an arbitrary modulation at 99\%~CL (Fig.~\ref{fig:LS_1}).

\section{Conclusion}
Despite apparent incompatibility with other experimental techniques using other targets, the persistence of the DAMA modulation claim continues to generate interest in NaI dark matter detectors. The approach we propose here would benefit from having particle identification, thanks to a simultaneous measurement of phonons and scintillation at cryogenic temperatures.  Currently the limiting factor would be the phonon channel, where attempts to date have encountered poor signal amplitude~\cite{Schaffner:2012ei}. Possible interpretations of this include the low Debye temperature of alkali halides and effects of hygroscopicity. Further work is necessary to understand if these problems can be overcome, for instance by using smaller individual detector modules. Other challenges to these detectors include the fragility of alkali halides, which may make them vulnerable to thermal contractions, and their hygroscopic nature that could restrict their handling to glove boxes and controlled atmospheres. These last two measures may be required in any case for other detectors as well to mitigate backgrounds.
In addition, work will be required to confirm the various nuclear recoil quenching factors at low temperature.

If these real technical challenges can be surmounted and phonon performance improved to the level of other materials, a short exposure of 10~kg-days would have a good chance of being able to fully explore the parameter space covered by the DAMA claim under standard astrophysical assumptions in the case of a time-independent analysis. \footnote{There remain a few scenarios our cryogenic detectors could not investigate, including that of dark atoms that bind to targets in a temperature-dependent manner~\cite{khlopov:2014}.} 
Furthermore, if a DAMA-type modulation exists, a modest detector array of 5~kg would have more than an 80\% chance of confirming it at a greater than 99\%~CL in less than 2~years if 99\% background rejection is achieved.
Indeed, an even smaller exposure may suffice as the DM-Ice~\cite{Cherwinka:2014xta} and SABRE~\cite{Xu:2014} collaborations are developing NaI(Tl) crystals with backgrounds potentially five times lower than DAMA. 
Lastly, pulse shape discrimination is already used in CsI at room temperature~\cite{kudryavtsev_csitl_2001,Kim:2012rza}. The large increase in light yield of CsI indicates it may be interesting to study pulse-shape discrimination as a form of background rejection at low temperatures.

\section{Acknowledgements} 
We thank V.~B.~Mikhailik for information on low-temperature  scintillation properties of alkali halides,  S.~Yellin for feedback on the statistics in the modulation section, and
W.~Rau and G. Gerbier for insightful comments on this manuscript. 
This work has been funded in Canada by NSERC (Grant SAPIN 386432), CFI-LOF and ORF-SIF (project 24536).

\section{References}
\bibliographystyle{model1-num-names}
\bibliography{alkhalsens}

\end{document}